\documentclass[conference]{IEEEtran}
\usepackage{mathrsfs}
\usepackage{algorithmicx}
\usepackage[ruled]{algorithm}
\usepackage{algpseudocode}
\usepackage{amssymb}
\usepackage{booktabs}
\usepackage{graphicx}

\usepackage{epsfig}
\usepackage{amssymb}
\usepackage{amsmath}
\usepackage{amsfonts}
\usepackage{booktabs}
\usepackage{verbatim}
\usepackage{graphicx}
\usepackage{hyperref}
\usepackage{subcaption}

\usepackage{multirow}
\usepackage{epstopdf}
\usepackage{multirow}

\usepackage{xcolor}
\usepackage{mathtools}
\let\labelindent\relax
\usepackage{enumitem}

\renewcommand{\algorithmicrequire}{\textbf{Input:}}
\renewcommand{\algorithmicensure}{\textbf{Output:}}

\algdef{SE}[SUBALG]{Indent}{EndIndent}{}{\algorithmicend\ }%
\algtext*{Indent}
\algtext*{EndIndent}

\begin{document}
%
\title{Criticality Aware Soft Error Mitigation in the Configuration Memory of SRAM based FPGA \vspace{-30pt}}
\author{\IEEEauthorblockN{Swagata Mandal\IEEEauthorrefmark{1},
Sreetama Sarkar\IEEEauthorrefmark{1},
Wong Ming Ming\IEEEauthorrefmark{1}, 
Anupam Chattopadhyay\IEEEauthorrefmark{1} and
Amlan Chakrabarti\IEEEauthorrefmark{2}}
\IEEEauthorblockA{Nanyang Technical University, Singapore\IEEEauthorrefmark{1}, University of Calcutta, Kolkata, India\IEEEauthorrefmark{2}}}


	


\maketitle

\begin{abstract}
Efficient low complexity error correcting code (ECC) is considered as an effective technique for mitigation of multi-bit upset (MBU) in the configuration memory (CM) of static random access memory (SRAM) based Field Programmable Gate Array (FPGA) devices. Traditional multi-bit ECCs have large overhead and complex decoding circuit to correct adjacent multi-bit error. 
In this work, we propose a simple multi-bit ECC which uses Secure Hash Algorithm for error detection and parity based two dimensional Erasure Product Code for error correction. Present error mitigation techniques perform error correction in the CM without considering the criticality or the execution period of the tasks allocated in different portion of CM. In most of the cases, error correction is not done in the right instant, which sometimes either suspends normal system operation or wastes hardware resources for less critical tasks. In this paper, we advocate for a dynamic priority-based hardware scheduling algorithm which chooses the tasks for error correction based on their area, execution period and criticality. The proposed method has been validated in terms of overhead due to redundant bits, error correction time and system reliability.   

\end{abstract}

\section{Introduction} \label{sec: Introduction}
\noindent With the increase in usage of SRAM based FPGA for various data intensive processes in different mission critical applications, soft error rate in the CM of FPGA devices increase due to radiated charged particles. Recent literature studies reveal that demand of FPGA devices has increased manifold nowadays compared to Application specific integrated circuit (ASIC) due to their low Non-Recurring Engineering cost, on field programmability, inherent parallelism~\cite{AMARA2006669}. Typically FPGA devices can be classified into three broad categories based on the technologies~\cite{7086415} used to store configuration data: Anti-fuse, flash and SRAM based FPGA. Till now anti-fuse based FPGAs are preferred for high radiation application, but one time programmable fuses within it prevent the users to change the configuration file once it is configured. On the other hand single event latch up, total ionizing dose~\cite{4033191} and limitation on the number of reconfiguration, debarred Flash based FPGA for long term missions. Commercially available SRAM based FPGA devices are quite flexible and have huge logic resources which are desirable for high performance computing. As most of the memory bits contain configuration data in SRAM FPGA~\cite{XYZ}, there is a high probability that configuration data will be corrupted due to radiation.
\par
Soft errors are temporary malfunctions that occur in solid state devices due to radiation. They are not reproducible~\cite{1197722} and sometimes lead to Single bit upset (SBU) and MBU in different embedded devices like FPGA. One of the common solutions to prevent FPGA devices from the effect of radiation is to use radiation hardened (Radhard) FPGAs like space grade FPGAs, but they are costlier compared to the commercial-off-the-self (COTS)~\cite{1369494} FPGAs and are also few generation behind COTS FPGAs. Hence, for different commercial applications, COTS FPGAs are used with various error mitigation techniques like triple modular redundancy or concurrent error detection~\cite{7019254} but they consume large area, huge power and are not suitable for real time applications. The problem related to extra overhead can be reduced partially by using scrubbing~\cite{6927476} where CM of FPGA devices is refreshed in a periodic interval with the stored configuration data (golden copy) in a separate Radhard memory. Though it reduces the effect of accumulated error and increases life span of FPGA devices, this method has to continuously access external Radhard memory, which increases the cost and introduces delay.
\par
Problem of storing the golden copy can be solved by using different error detection and correction (EDAC) codes like Bose, Chaudhuri, Hocquenghem (BCH)~\cite{1046102} code for error mitigation in configuration data but their decoding complexity and latency are quite high. In general, soft errors are localized in nature i.e it corrupts adjacent multiple bits~\cite{MANDAL2017313} and hence, complicated EDAC with high redundancies are required to efficiently correct the effect of adjacent erroneous bits. Sometimes concatenated code or product code may be good solutions for mitigation of adjacent MBU (AMBU) as it corrects data along both row and column of storage element in parallel. Since error detection can be done more easily compared to error correction~\cite{7104165}, here we have separated error detection from correction and proposed error detection methodology in the CM using secure hash algorithm (SHA)~\cite{keccak1}. Though SHA is used to test the data integrity in cryptography, we have used it to detect the presence of erroneous bits in the configuration data. After error detection, simple parity based erasure product code will be used to correct AMBU in the CM.
\par
\begin{figure}[ht]
\vspace{-8pt}	
\centering
\includegraphics[scale=0.28]{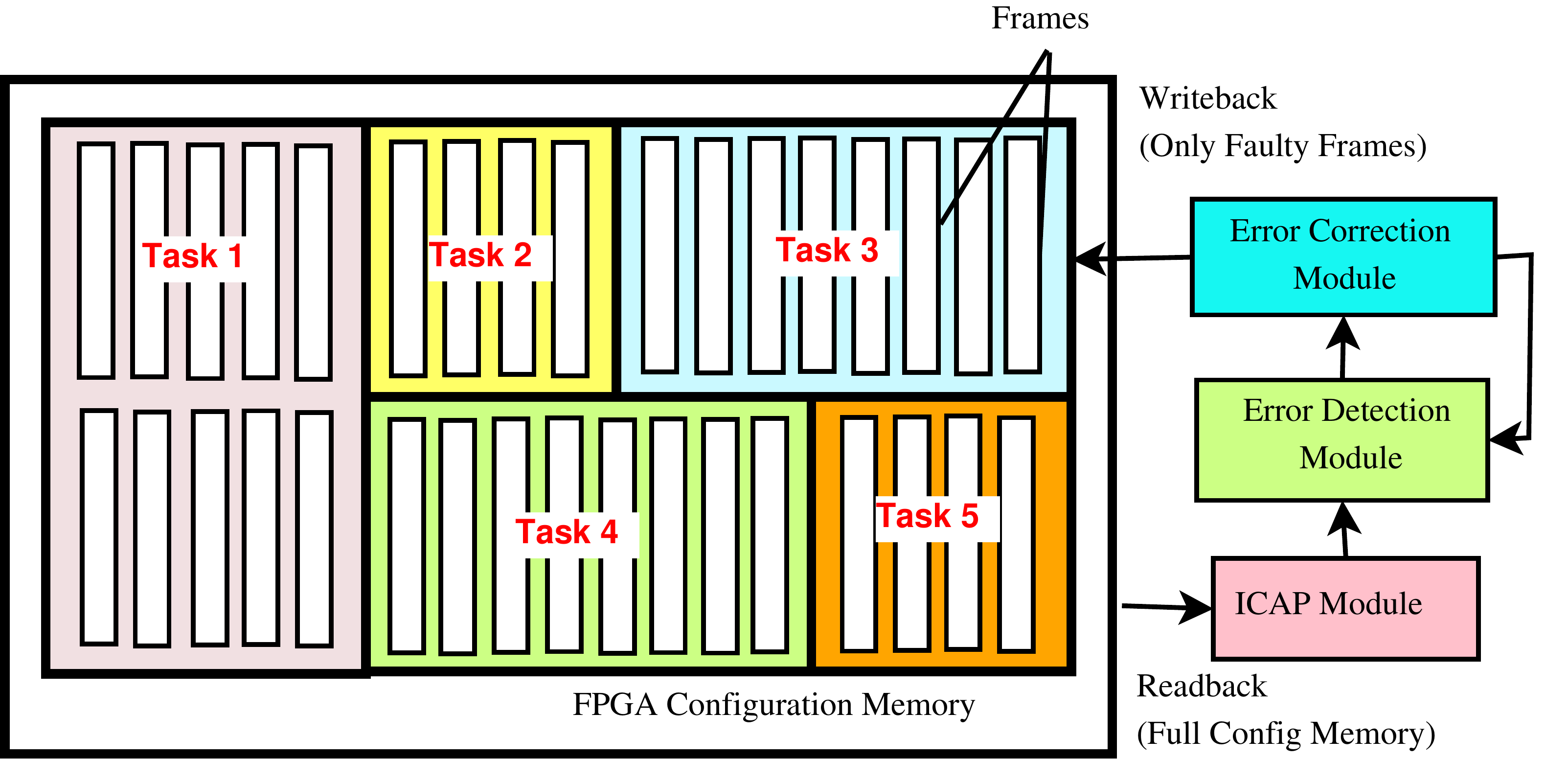}
\vspace{-8pt}
\caption{\em Error detection and correction flow in the CM of FPGA}
\label{fpgaconfig}
\vspace{-20pt}
\end{figure}
During the design of the complex FPGA based system, CM is partitioned into multiple partial reconfiguration regions (PRRs) and different tasks are allocated on each PRR as shown in Figure~\ref{fpgaconfig}. Error correction in CM involves read back of data from PRR, error detection and correction on this data and downloading it into CM. In most of the cases priorities of different tasks allocated to PRR regions are not considered, which may create problem for different real time applications.
In order to alleviate this problem, we have proposed dynamic hardware scheduling algorithm that downloads PRR based on the criticality and area of the task allocated on the PRR without suspending normal system operation. In this paper our key contributions are:
\vspace{-8pt}
\begin{itemize}
[leftmargin=0.3cm]
\item{An efficient soft error mitigation method is proposed which separate error detection from error correction. Area optimized SHA-3 for error detection and erasure product code for error correction are proved to be quite efficient compared to other state of art solutions.}

\item{A hardware scheduling algorithm is proposed which calculate the priority for the reconfiguration of the tasks based on criticality, area and execution period of the tasks.}  
\end{itemize}
The rest of the paper is organized as follows. Section~\ref{sec:prelim} presents a detailed literature review related to our work, section~\ref{sec:method}describes the proposed error detection and correction method in details and section~\ref{sec:hsatest} illustrates hardware scheduling scheduling algorithm. Performance evaluation with result analysis is described in section~\ref{sec:perresult} followed by concluding remarks in section~\ref{sec:conc}.

\vspace{-6pt}
\section{Literature Review}\label{sec:prelim}
\noindent With the development of fabrication technology, solid state devices are gradually reducing in size, hence node voltage of CMOS transistor also reduces. This increases the probability of occurrence of AMBU in the CM of FPGA devices.  In order to reduce error correction complexity, designers always prefer simple EDAC code with high error correcting capability and less redundancy to correct multi-bit error. In general, Xilinx provides soft error mitigation controller using single bit error correcting Hamming code for this purpose  but this is not sufficient to correct MBU in the CM. Authors in~\cite{5682391} proposed two dimensional Hamming product code to correct multi-bit error in the CM of SRAM based FPGA though the proposed method is unable to correct the erroneous bits when multiple erroneous bits are present along both rows and columns of the memory element. Hamming code is concatenated with parity code or BCH code in~\cite{6187474}  for correction of multiple bits in the memory element.
\par
In order to make the error correction simpler, authors in ~\cite{MANDAL2017313} separated error detection from correction where parity code with interleaving is used to detect erroneous configuration frames, though detection efficiency is varied with the interleaving depth. In this paper, we propose 512 bit SHA-3~\cite{keccak1} function to detect the presence of erroneous bits in the configuration data. SHA function is widely deployed in digital signature schemes, message authentication codes (MACs) and several other information security applications. The most essential properties of a secure cryptographic hash function are being simple in computation and yet highly non-invertibile with strong collision resistance. In cryptography, bit flip in a data stream due to collision is equivalent to bit flip due to radiation in the configuration data. Though hardware architecture of SHA-3 is slightly complex compared to parity based error detection module, its error detection capability is always 100\%.
\par
Downloading of partial bit file for a task in the CM sometimes affects the functionalities of other tasks, so proper hardware scheduling is needed along with scrubbing or error correction of configuration data. Authors in~\cite{6927476} proposed a criticality aware scheduling algorithm which scrub different PRR in FPGA, based on criticality of the task allocated to the PRR. The main problem of this method is that criticality of different tasks are fixed and scrubbing sequence of different tasks remain same. If any task is stopped or any new task is initiated, scrubbing sequence will not change i.e they do not support run time adaptation. In order to solve this problem authors proposed dynamically adaptive scrubbing technique in~\cite{7167341} which improves reconfiguration process in FPGA. As there is only one downloading port is available in the Internal Configuration Access Port (ICAP) proper port scheduling is also necessary to download $n$ tasks in $m$ PRR (n$\geq$m) as described by authors in~\cite{4211783} for hard real time reconfiguration system. Authors proposed a method which integrates error detection and correction with dynamic priority based hardware scheduling in~\cite{MANDAL2017313} but here tasks are periodic in nature and criticality of the tasks are not included. Unlike to the previous case, here our proposed soft error mitigation for aperiodic tasks calculates the priority for reconfiguration of tasks considering its criticality, execution time and area.  

\vspace{-6pt}
\section{Proposed EDAC method}\label{sec:method}
\noindent 
In the proposed method we have separated error correction from detection.
Here error is detected using 512 bit SHA-3 and erasure product code is used for error correction. SHA-3 is a new member of secure hash algorithm family and it's architecture is different from SHA-1 and SHA-2.
\vspace{-10pt}
\subsection{Error detection using SHA-3}
SHA produces fixed length digital signature when a variable length data stream is applied at its input. Change in any single bit in the input data stream will change the digital signature randomly. We have used this property of SHA for error detection in configuration data. The configuration data should remain unchanged after configuration of FPGA devices to ensure fault-free operation. Before downloading the bit file in CM, digital signature for each task is stored in the flash memory. During error detection, configuration data will be read back and configuration data for each task will be passed through SHA-3 module to produce its digital signature.  Presence of error will be confirmed in a particular task if there is a mismatch between the stored and the newly computed digital signatures. 
\par
\begin{figure}[ht]
\vspace{-20pt}	
\centering
\includegraphics[scale=0.45]{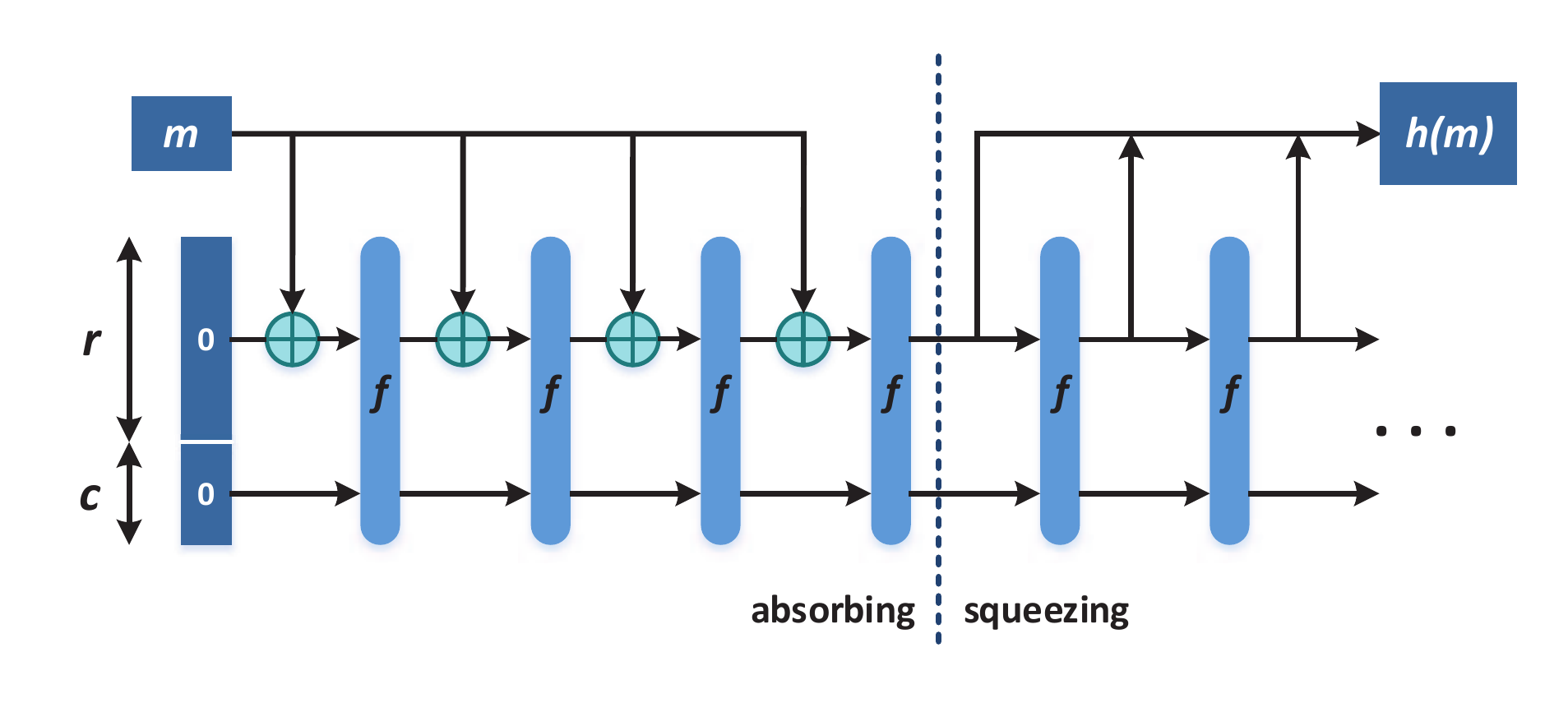}
\vspace{-20pt}
\caption{\em Keccak sponge construction}
\label{Kesponge}
\vspace{-26pt}
\end{figure}

The Keccak hash function \cite{keccak1}, designed by G. Bertoni et. al., was announced by the NIST as the new Secure Hash Algorithm-3, in 2015. Generally, the Keccak algorithm is based on sponge construction, where the hash transformation is performed on an internal state that takes input of arbitrary length, and produces an output of the desired length. SHA-3 algorithm consists of two phases: absorbing and squeezing phase as shown in Figure~\ref{Kesponge}. Each state in the sponge function consists of bitrate ($r$) and capacity($c$). In the absorbing phase, the bit rate of the initialized state is XORed with the first part of the input. The new bitrate, together with the capacity of the initialized state matrix, will form a new state that is used in f-permutation. 
The resulting state will serve as the new initial state for the next round and the process continues for 24 iteration rounds. Each round is divided into five separate steps, i.e. Theta ($\theta$), Rho ($\rho$), Pi ($\pi$), Chi ($\chi$) and Iota ($\iota$). In the squeezing phase, first $r$ bits of the internal state contains the final output.
\par
\begin{figure}[ht]
\vspace{-10pt}	
\centering
\includegraphics[scale=0.6]{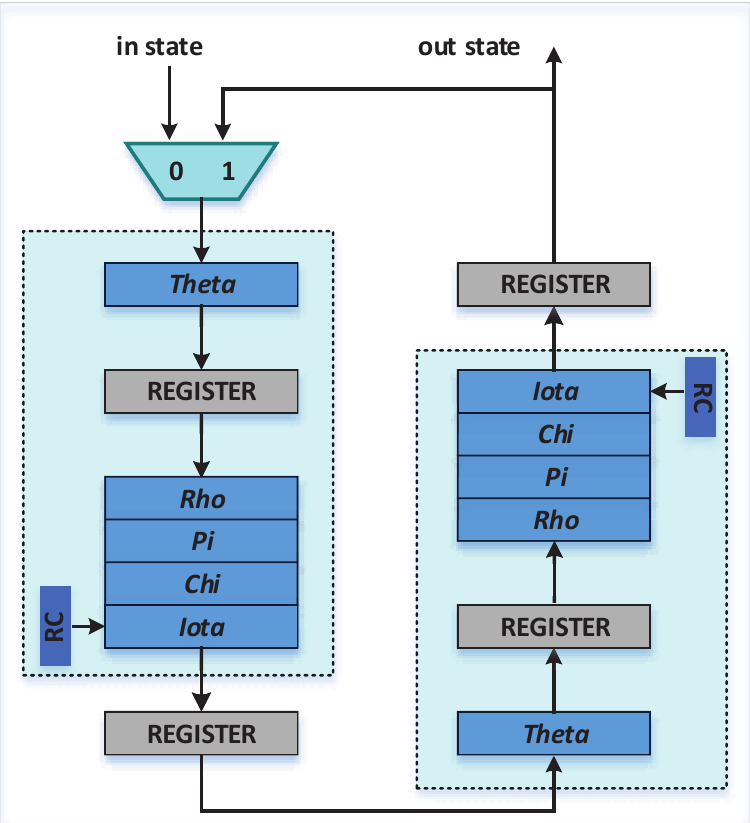}
\caption{\em Architecture of optimized SHA-3 Algorithm}
\label{fig:subpipe}
\vspace{-20pt}
\end{figure}

In this study, we have chosen to use $512$-bits Keccak[$1600$] due to its guaranteed security margin. Therefore, the respective value of $r$ and $c$ are $576$ and $1024$. With this, $1600$-bit state matrix of Keccak is composed of $5 \times 5$ matrix of $64$-bit words. The proposed modified SHA-3 architecture utilizes the concept of unrolling, pipelining and subpipelining~\cite{8351649} as depicted in Figure \ref{fig:subpipe}. Features of the proposed optimizations are:
\vspace{-5pt}
\begin{itemize}
\item Simplified round constant generator (storing only the non-zero bits) 
\item $2$-stages sub-pipelining within transformation round (inserted after the Theta ($\theta$))
\item Unrolling factor of $2$   
\item $2$-stages pipelining in between adjacent rounds.
\end{itemize}
It is worth a note that as a result of the subpipelining, the longest delay in the first half of the computation is constituted of $5$ XORs. Meanwhile, the second part which includes Pi ($\pi$) to Iota ($\iota$) covers the longest delay of $2$ XORs, $1$ AND and $1$ XOR. Users can consult~\cite{8351649} for details of the proposed architecture.
\vspace{-10pt}
\subsection{Error Correction using Erasure code}
\begin{figure}[ht]
\vspace{-11pt}	
\centering
\includegraphics[scale=0.45]{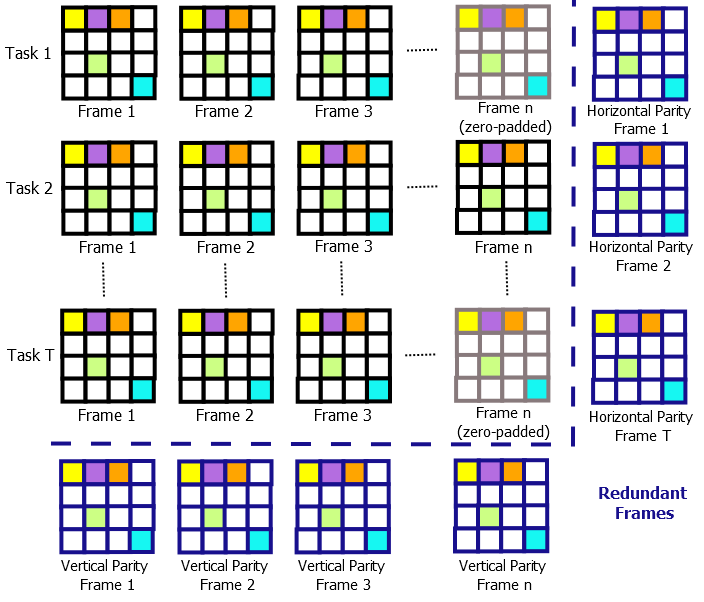}
\caption{\em Error correction using Ersaure product code}
\label{figerasure}
\vspace{-5pt}
\end{figure}

This subsection describes the proposed error correction algorithm. Erasure code is an error correcting method which converts $m$ blocks of data into $n$ blocks of data in such a way that it can recover any erased data block from $m+n$ data blocks~\cite{MANDAL2017313}. Here we have used two dimensional Erasure product code which can recover any number of erroneous bits in any single configuration frame of a task. Erasure product code is basically a parity based coding which can correct any erroneous data bits in a memory element with the help of both vertical and horizontal parity bits. As error correction is performed along both row and column of the memory element in parallel, decoding is very fast and simple. 
The initial part of decoding and encoding are quite similar as both are involved in the parity calculation.  As shown in Figure~\ref{figerasure}, the configuration frames are arranged in the form of a two dimensional array, where each row consists of the frames of a task. To compensate for the varying number of frames in different tasks, some dummy frames are added whose each bit contains zero value. In Figure~\ref{figerasure}, bits marked with the same color will be XORed to generate horizontal and vertical parity bits.
Before downloading the configuration data, the hashes for the different tasks are calculated using 512 bit SHA-3 function illustrated in previous subsection and stored in the flash memory. Also, horizontal and vertical parity of the original configuration frames allocated for different tasks will be calculated  and stored.
\par
Decoding process starts with reading of the entire data from the CM of FPGA. The pseudocode for the proposed technique is given in algorithm~\ref{errocorralgo}. In the proposed algorithm, number of tasks, number of frames in a task, total number of frames, number of column and row in each frame are represented by $N$, $n$, $U$, $h$ and $v$ respectively. `hash[N][512]' store 512 bit hash for N tasks.
\begin{algorithm}
\caption{Error correction algorithm}
\label{errocorralgo}
\begin{algorithmic}[1]
\Require $n$, $N$, $h$, $v$, $U$;
\Ensure task\_correct;
\For{runs = 1 $\to$ no\_of\_runs}
\State{inject\_error();}
\State{Calculate horizontal parity frame from erroneous tasks and store it into hpf[N][v][h];}
\State{Calculate vertical parity frame from erroneous tasks and store it into vpf[n][v][h];}
\For{$z = 1 \to N$}
\State{task\_correct = 0;}
\For{$j = 1 \to v$}
\For{$i = 1 \to h$}
\If{(hp[z][j][i] $\neq$ hpf[z][j][i])}
\State{in = 0;}
\For{$k = 1 \to n$}
\If{(vp[k][j][i] $\neq$ vpf[k][j][i])}
\State{choose\_frame[in] = k;}
\State{in = in + 1;}
\EndIf
\EndFor
\State{task\_correct = shc(z,choose\_frame,in);}
\EndIf
\If{(task\_correct = 1)}
\State{Break;}
\EndIf
\EndFor
\If{(task\_correct = 1)}
\State{Break;}
\EndIf
\EndFor
\EndFor
\EndFor
\State{\textbf{def} shc (z,choose\_frame,in)}
\State{task\_correct = 0;}
\For {x = 1 $\to$ in}
\State{frame\_no = choose\_frame[x];}
\State{err\_frame = ConF(z*n+frame\_no);}
\For{$j = 1 \to v$}
\For{$i = 1 \to h$}
\If{(hp[z][j][i] $\neq$ hpf[z][j][i])}
\State{err\_frame[j][i] = err\_frame[j][i]$\oplus$1;}
\EndIf
\EndFor
\EndFor
\If{(hash[z] = sha512(task[z]))}
\State{task\_correct = 1; Break;}
\Else
\For{$j = 1 \to v$}
\For{$i = 1 \to h$}
\If{(hp[z][j][i]$\neq$ hpf[z][j][i])}
\State{err\_frame[j][i] = err\_frame[j][i] $\oplus$ 1;}
\EndIf
\EndFor
\EndFor
\EndIf
\EndFor
\State{\textbf{return} task\_correct}
\Statex
\end{algorithmic}
\vspace{-12pt}
\end{algorithm}
The inject\_error function injects a random error pattern into the CM of FPGA. The horizontal and vertical parity is again calculated after error infusion. If coordinate (j,i) of task z does not match with the original parity bit of the respective horizontal parity frame, the vertical parity frames are checked for the same coordinates, and the frames in which anomaly is recorded, are stored in the choose\_frame array and passed onto the $shc$ function. A frame in the choose\_frame array is first corrected by comparing it with the original horizontal parity frame for that task and matched with the stored hash value for that task. If it matches, we successfully obtain the faulty frame in that task, but if it does not, the frame is reverted back to its original position and we try the same method with the other frames in the choose\_frame array until the faulty frame is corrected.

\vspace{-10pt}
\section{Hardware Scheduling Algorithm}\label{sec:hsatest}
\noindent 
Configuration memory of FPGA devices are partitioned into multiple $PRR$ as shown in Figure~\ref{fpgaconfig} and each $PRR$ is assigned an individual task. Task allocated to each $PRR$ is either independent or dependent on other tasks and consists of multiple configuration frames. Lets us assume in our application, available $N$ number of tasks execute aperiodically and out of these errors are detected in $L$ number of tasks. 
\begin{algorithm}
\caption{Algorithm of download manager}
\label{scheddown}
\begin{algorithmic}[1]
\Require $w_a$, $w_b$, $w_c$, $busy_i$, $T_i$ where $i = 1\to L$;
\Ensure $FP_{max}$;
\State{Form Task Dependency Graph for $\Gamma_i$ where $i = 1\to N$;}
\State{Calculate criticality of each Task for $\Gamma_i$ where $i = 1\to L$;}
\State{Store criticality of Task $\Gamma_i$ in an array $\zeta$, $\forall$  $i = 1\to L$;}
\For{$i = 1 \to L$}
\If{$(busy_i$ =`1')}
\State{$St_i=(\frac{E_i}{t}-PE_i)+\frac{I_i}{t}$}
\EndIf
\If{$(busy_i$ =`0')}
\State{$St_i=\frac{I_i}{t}-PI_i$}
\EndIf
\If{(rising\_edge($clk$))}
\If{$(St_i=0)$}
\State{$St_i=\frac{E_i+I_i}{t}$}
\Else
\State{$St_i=St_i-1$;}
\EndIf
\If{($RT_i+EC_i<=St_i * t$)}
\State{$P_i=St_i-$($\frac{EC_i +RT_i}{t}$)}
\EndIf
\EndIf
\EndFor
\State{$FP_i=$[$w_a\times (\frac{1}{P_i})+w_b\times (\frac{\eta_i}{\eta})+w_c\times \zeta(i)+w_d\frac{E_i}{t}$]}
\State{Find maximum among all$FP_i$;}
\Statex
\end{algorithmic}
\end{algorithm}
\setlength{\textfloatsep}{0pt}
The error detection time, error correction time, reconfiguration time, execution time and idle time of $i^{th}$ task are designated as $ED_i$, $EC_i$, $RT_i$, $E_i$ and $I_i$ respectively. $t$ is the time period of the clock which drives the system. $w_a$, $w_b$, $w_c$ and $w_d$ are user defined parameters. Our proposed hardware scheduling algorithm~\ref{scheddown} will download the $L$ tasks in the CM without hampering the normal functionalities of other tasks. 
\vspace{-5pt}
\par
The first step of our proposed algorithm is the calculation of criticality of each task. Here the criticality of a task measure dependency of a task on other tasks. Criticality of a task can be defined as the ratio of the number of tasks dependent on that task to the number of tasks present in the system.
\begin{figure}[ht]
\vspace{-8pt}	
\centering
\includegraphics[scale=0.35]{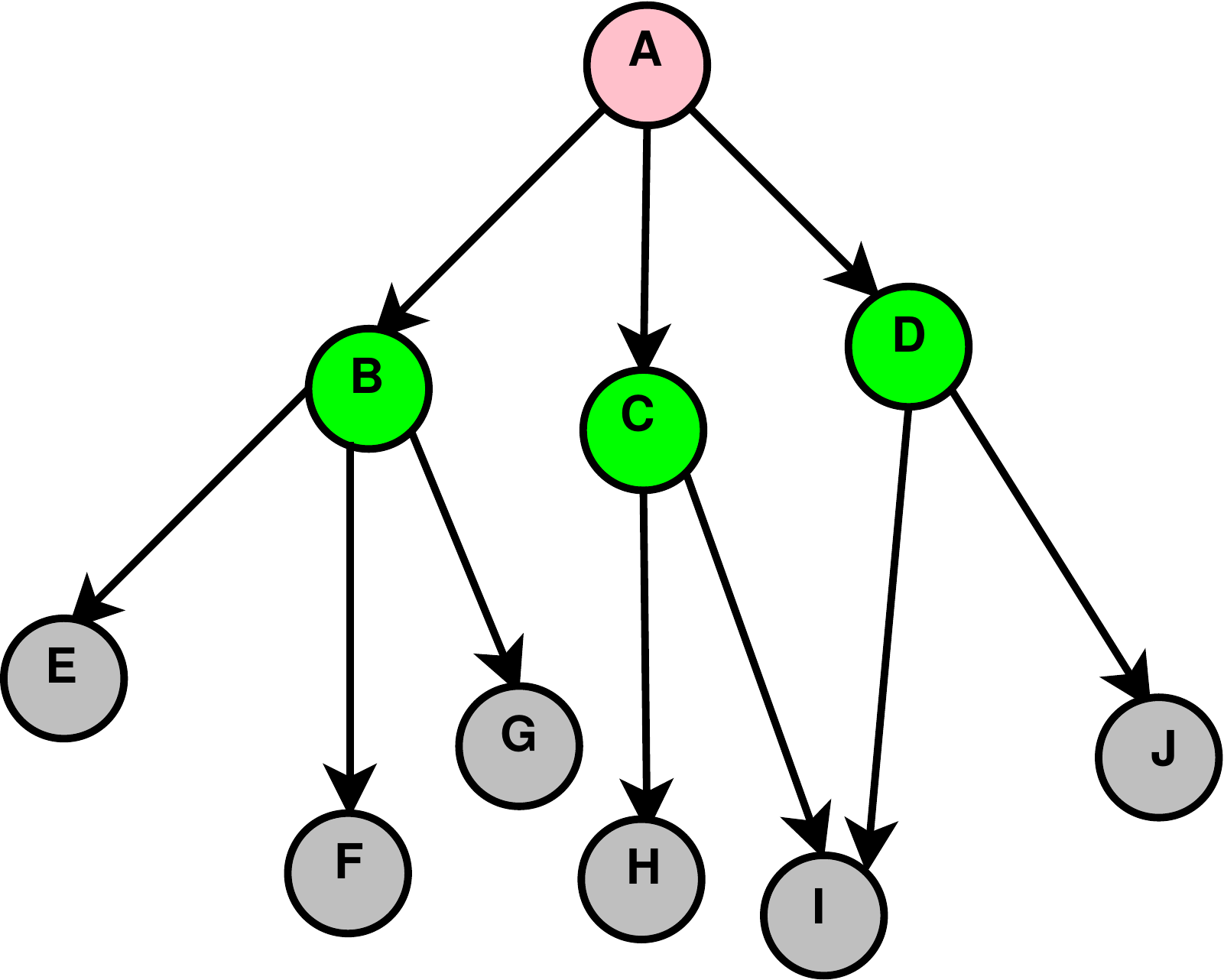}
\caption{\em Example of task dependency graph}
\label{tsgfpga}
\vspace{-20pt}
\end{figure}
 In order to calculate criticality of a task, we take the help of task dependency graph~\cite{Satish1450079}. Figure~\ref{tsgfpga} shows a task dependency graph assuming ten tasks are present in the CM. The task on which more number of other tasks are dependent, is more critical compared to other tasks. As for example, all other tasks depend on task A, so criticality of task A will be 0.9. Similarly criticality of task B, C and D will be 0.3, 0.2 and 0.2 respectively. Criticality of all other tasks will be 0 because no other tasks depend on them. Here we assume that task dependency graph is a directed acyclic graph i.e that is there is no cyclic dependency among the tasks because presence of cyclic dependency on different tasks will complicate criticality calculation. Criticality of only erroneous tasks will be stored in an array $\zeta$ because finally they will be used in priority calculation. Steps of the proposed algorithm can be described as follows:
\vspace{-6pt}
\begin{itemize}
\item{During downloading of configuration file of a PRR, task allocated to that PRR must be remain idle.}
\item{Each task is associated with three signals: busy, partial execution (PE), partial idle (PI). Busy signal will be high during the execution phase of the task and remain low during idle phase. PE and PI count the number of clock cycles after starting of the execution phase and idle phase of a task respectively. A status register $St_i$ associated with each task measure the number of clock cycles from the current time to the initiation of next execution phase of the task i.e., $St_i$ will be loaded with slack time. At each rising edge of clock, $St_i$ will be decremented by 1 and when it will be 0, it will be loaded by $\frac{E_i+I_i}{t}$.}
\item{In the next step priority $P_i$ will be calculated by subtracting $\frac{EC_i+RT_i}{t}$ from $St_i$.}
\item{Final priority ($FP_i$) will depend on additional parameters alongside $P_i$ like execution time, number of configuration frames and criticality of the task. The task which has more number of configuration frames has more chance to be affected by MBU which is reflected in $FP_i$ as the ratio of number of configuration frames in a task ($\eta_i$) to the total configuration frames ($\eta$) in the CM. The erroneous task with longer execution time will give erroneous result for a longer time compared to the faulty task with smaller execution time. Hence, user will first try to correct a faulty task with longer execution time. Similarly, task with higher criticality provides erroneous results to more number of dependent tasks compared to tasks with lower criticality, so it is always advisable that faulty task with higher criticality be corrected as early as possible. Now based on the user input ($w_a$, $w_b$, $w_c$ and $w_d$) $FP_i$ will be calculated as described in algorithm~\ref{scheddown}. Values of user inputs can varied in between 0 and 1. Here $P_i$ and $FP_i$ will be updated in parallel.}
\item{During error correction of a faulty task, $FP_i$ will be monitored for the remaining ($L-1$) tasks and again the task with highest $FP_i$ will be chosen for error correction.}
\item{When $FP_i$ for multiple tasks will be same, scheduler will download the task whose $St_i$ is smaller, i.e., scheduler will follow early deadline fast (EDF) algorithm~\cite{6927476}.}
\end{itemize}

\vspace{-10pt}
\section{Performance Analysis}\label{sec:perresult}
\begin{figure}[ht]	
\centering
\includegraphics[scale=0.3]{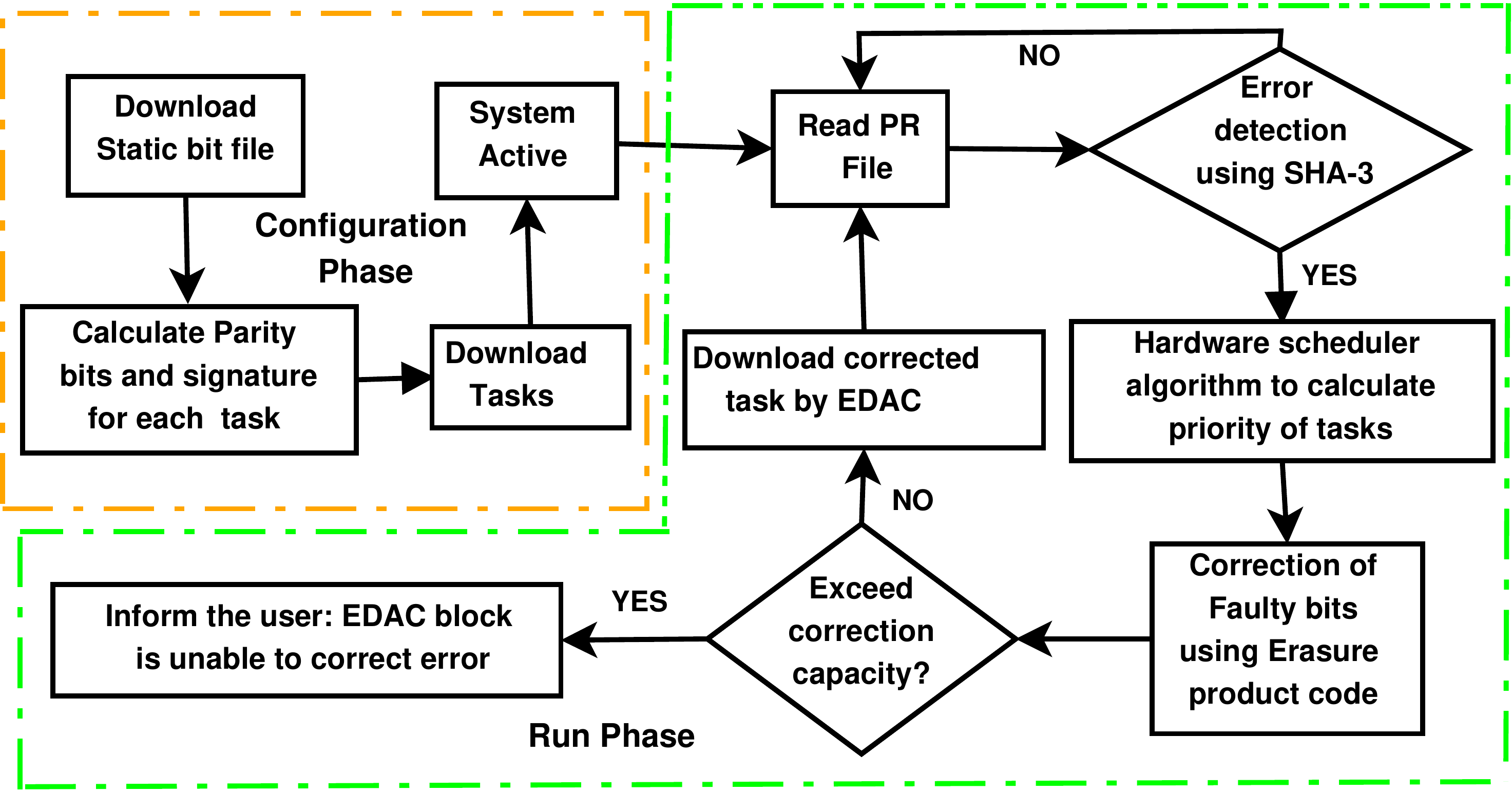}
\caption{\em Workflow of proposed error correcting model}
\label{flowcheck}
\vspace{-5pt}
\end{figure}

Proposed error detection and correction methods have been implemented on the Xilinx Kintex7 board using Vivado platform and VHDL for design entry. We have tested our design using behavioral simulations. To validate EDAC capability of the proposed method, we have performed fault injection experiment in the CM of FPGA. The proposed design flow is shown in figure~\ref{flowcheck} which consists of configuration and run phase. The whole CM is partitioned into two parts: static part which contains EDAC module and dynamic part which contains PRR. 
During configuration phase, bit file of EDAC module and the tasks will be downloaded into the CM and parity bits and signature for each task will be calculated. During run phase, bit file corresponding to PRR will be read back and passed through error detection module. If error is detected, the task scheduler will calculate the priority of faulty tasks and erasure code will correct faulty tasks. Here error detection using SHA-3 is always 100\% but correction using erasure product code is only possible if the erroneous bits are present within a configuration frame of a task. In general, tasks are placed physically in different locations of CM so there is a high probability that single configuration frame of a task will be affected by soft error. Hence, here average error correction probability is almost 100\%. If multiple frames of a tasks is affected by radiation system, EDAC block will inform the user that error is present in the task but correction is not possible.
\begin{figure}[ht]	
\vspace{-10pt}
\centering
\includegraphics[scale=0.5]{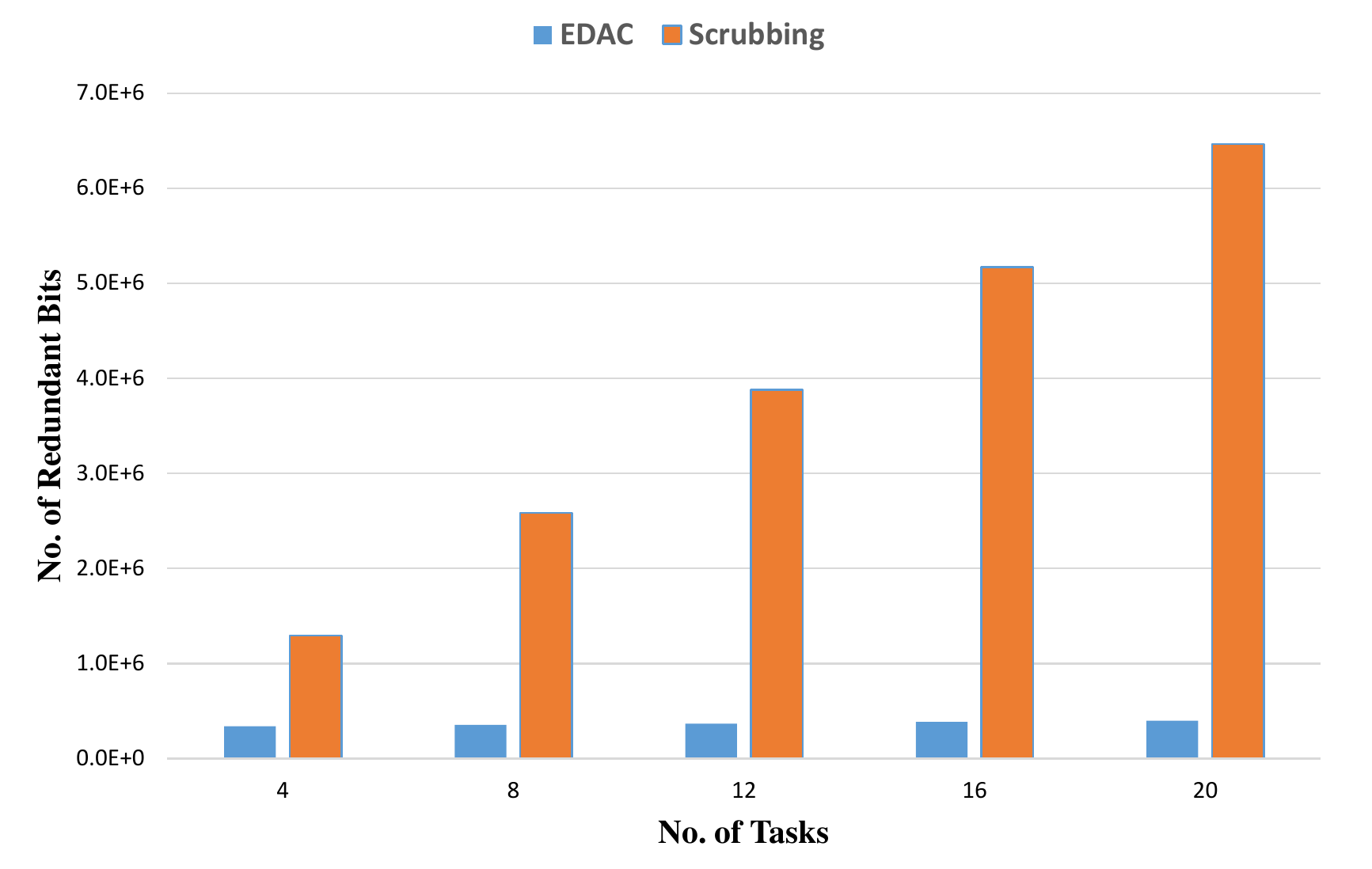}
\caption{\em Comparison of redundant bits of our proposed method and scrubbing proposed by authors in~\cite{6927476}}
\label{reduncom}
\vspace{-25pt}
\end{figure}
\par
Figure~\ref{reduncom} compares the redundant bits required for both error detection and correction using our proposed model and scrubbing model proposed in~\cite{6927476}. In scrubbing, all the configuration frames need to be stored but in our case signature for each task and parity bits for all tasks will be stored. This will reduce storage memory requirement drastically as illustrated in figure~\ref{reduncom}. 
\par
Table~\ref{table:comparison} investigates the performance of our proposed SHA-3 architecture in terms of the area, throughput and efficiency and compares it with the state of the art solutions. Here throughput is calculated using the equation~\ref{equ1}.
\begin{equation}\label{equ1}
\vspace{-5pt}
Throughput = \frac{\# blocksize \times F_{max}}{\#clockcycle}\times N_{msg}
\end{equation}
where $\#blocksize$ refers to the number of processed bits, $\#clockcycle$ corresponds to the total clock cycles between
successive messages to generate each digest message and $\#F_{max}$ is the highest attainable frequency in the implementation.$\#N_{msg}$ is the number of messages that can be simultaneously hashed at a given time. Based on these investigations, subpipelining is observed effective in critical path reduction while unrolling with pipelining enables simultaneous processing in SHA-3 hash function. Generally, both of the attributes bring positive effect on the throughput performance. 
\begin{table}
\begin{center}
\caption{Comparison of our proposed SHA-3 with other SHA-3 design in terms of area, throughput and efficiency}
\scalebox{0.8}{
\begin{tabular}{|c|c|c|c|c|c|}
\hline
Hash&\shortstack{Feature\\(Pipeline\/Unroll)}&\shortstack{Fmax\\MHz}&\shortstack{Area\\Slices}&\shortstack{Throughput\\Gbps}&\shortstack{Efficiency\\Mbps\/Slices}\\
\hline
\shortstack{Proposed\\Architecture}&\shortstack{Unrolled k=2\\ Pipeline n=2 \\ Subpipeline n=2}&344&1406&16.51&11.47\\
\hline
Athanasiou et al~\cite{8351649}& Subpipeline n=2 &397&1649&9.55&5.80\\
\hline
Ioannou et al~\cite{8351649}&\shortstack{Unrolled k=2\\ Pipeline n=2}&391&2296&18.77&8.17\\
\hline
Michail et al~\cite{8351649}&\shortstack{Unrolled k=3\\ Pipeline n=3}&391&3965&28.15&7.10\\
\hline
\end{tabular}}
\label{table:comparison}
\end{center}
\vspace{-20pt}
\end{table}
Table~\ref{table:reduneffi} compares error detection capability of the SHA-3 with other error detection methods proposed in~\cite{MANDAL2017313},~\cite{7104165} where authors have used parity with interleaving along different dimensions of CM for error detection. Here we have considered ten tasks in CM and each has hundred configuration frames. With the increase of redundant bits, error detection capability of the methods proposed in ~\cite{MANDAL2017313},~\cite{7104165} increase, whereas in our case error detection is always 100\% with very less number of redundant bits as illustrated in Table~\ref{table:reduneffi}. 
\begin{table}
\begin{center}
\caption{Comparison of error detection capability of SHA-3 with other error detection methods}
\scalebox{0.7}{
\begin{tabular}{|c|c|c|c|c|c|c|c|}
\hline
\multicolumn{2}{|c|}{I2D~\cite{7104165}} & \multicolumn{2}{|c|}{I3D~\cite{7104165}} & \multicolumn{2}{|c|}{IMMC~\cite{MANDAL2017313}} &\multicolumn{2}{|c|}{SHA-3}\\
\hline
Redundancy&Efficiency&Redundancy&Efficiency&Redundancy&Efficiency&Redundancy&Efficiency\\
(bits)&&(bits)&&(bits)&&(bits)&\\
\hline
3000&81.79\%&8000&93.55\%&8000&96.25\%&5120&100\%\\
7000&96.75\%&11000&97.9\%&16000&99.61\%&&\\
10000&97\%&13000&98.85\%&20000&100\%&&\\
\hline
\end{tabular}}
\label{table:reduneffi}
\end{center}
\vspace{-4pt}
\end{table}
\par
\begin{figure}[ht]	
\vspace{-10pt}
\centering
\includegraphics[scale=0.53]{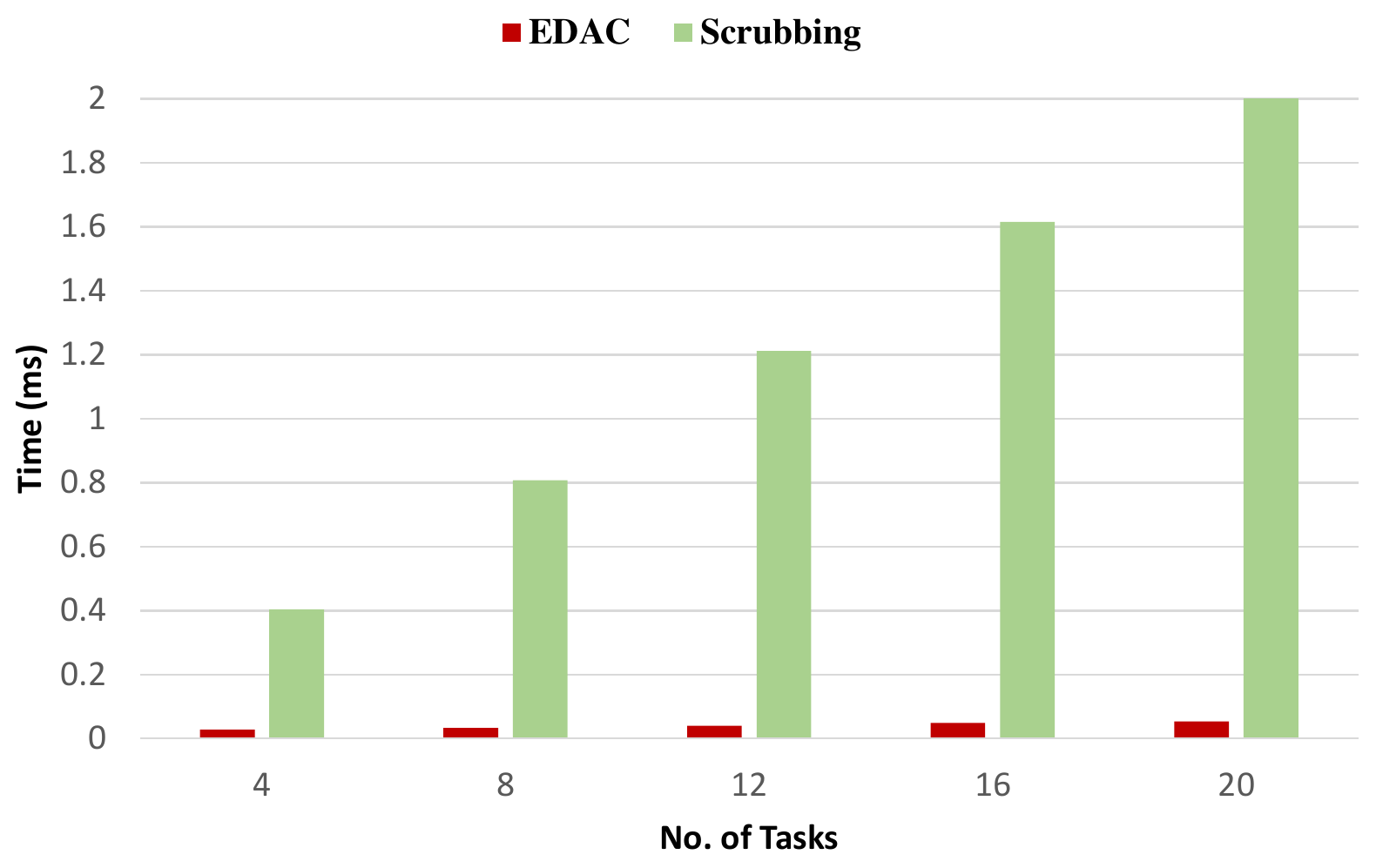}
\vspace{-5pt}
\caption{\em Comparision of Error detection and correction time of the proposed method with the scrubbing~\cite{6927476}}
\label{errdetcotime}
\vspace{-14pt}
\end{figure}

\vspace{-5pt}
During scrubbing, all configuration frames need to be downloaded. Though it eliminates the effect of error accumulation, it increases error correction time compared to our proposed model. This is due to the fact that time required for error correction and downloading of only erroneous configuration frames in the faulty tasks is very less compared to downloading of all configuration frames in scrubbing as shown in figure~\ref{errdetcotime}. 
In our scheduling model, we have proposed dynamic updation of the priority of the tasks whereas authors in~\cite{6927476} keep the criticality of the tasks fixed. We have also considered area and execution time of the tasks, which makes our proposed task scheduling more realistic and reliable.

\vspace{-10pt}
\section{Conclusion}\label{sec:conc}
In this work we have proposed an error correcting model which uses SHA-3 for error detection and simple parity based erasure code for mitigation of soft errors in the configuration memory of FPGA devices. Dynamic partial reconfiguration along with hardware scheduling algorithm which schedule the reconfiguration of different faulty tasks based on their criticality, area and execution time helps in error mitigation in the configuration memory without suspending normal system operation. Experimental results prove that
our proposed models deliver better performance in terms of error correction time and overhead compared to other state of the art solutions.
\vspace{-8pt}
\IEEEpeerreviewmaketitle
{\scriptsize 
\bibliographystyle{ieeetr}
\bibliography{IEEEexample}

\begin{thebibliography}{10}

\bibitem{AMARA2006669}
A.~Amara, F.~Amiel, and T.~Ea, ``Fpga vs. asic for low power applications,''
  {\em Microelectronics Journal}, vol.~37, no.~8, pp.~669 -- 677, 2006.

\bibitem{7086415}
M.~Wirthlin, ``High-reliability fpga-based systems: Space, high-energy physics,
  and beyond,'' {\em Proceedings of the IEEE}, vol.~103, pp.~379--389, March
  2015.

\bibitem{4033191}
H.~J. Barnaby, ``Total-ionizing-dose effects in modern cmos technologies,''
  {\em IEEE Transactions on Nuclear Science}, vol.~53, pp.~3103--3121, Dec
  2006.

\bibitem{XYZ}
Xilinx, ``Logicore ip soft error mitigation controller v3.4,product guide,''
  2012.

\bibitem{1197722}
H.~T. Nguyen and Y.~Yagil, ``A systematic approach to ser estimation and
  solutions,'' in {\em 2003 IEEE International Reliability Physics Symposium
  Proceedings, 2003. 41st Annual.}, pp.~60--70, March 2003.

\bibitem{1369494}
M.~Violante, ``Simulation-based analysis of seu effects in sram-based fpgas,''
  {\em IEEE Transactions on Nuclear Science}, vol.~51, pp.~3354--3359, Dec
  2004.

\bibitem{7019254}
I.~M. Safarulla and K.~Manilal, ``Design of soft error tolerance technique for
  fpga based soft core processors,'' in {\em 2014 IEEE International Conference
  on Advanced Communications, Control and Computing Technologies},
  pp.~1036--1040, May 2014.

\bibitem{6927476}
R.~Santos, S.~Venkataraman, A.~Das, and A.~Kumar, ``Criticality-aware scrubbing
  mechanism for sram-based fpgas,'' in {\em 2014 24th International Conference
  on Field Programmable Logic and Applications (FPL)}, pp.~1--8, Sept 2014.

\bibitem{1046102}
L.~Song, M.-L. Yu, and M.~S. Shaffer, ``10- and 40-gb/s forward error
  correction devices for optical communications,'' {\em IEEE Journal of
  Solid-State Circuits}, vol.~37, pp.~1565--1573, Nov 2002.

\bibitem{MANDAL2017313}
S.~Mandal, R.~Paul, S.~Sau, A.~Chakrabarti, and S.~Chattopadhyay, ``Efficient
  dynamic priority based soft error mitigation techniques for configuration
  memory of fpga hardware,'' {\em Microprocessors and Microsystems}, vol.~51,
  pp.~313 -- 330, 2017.

\bibitem{7104165}
M.~Ebrahimi, P.~M.~B. Rao, R.~Seyyedi, and M.~B. Tahoori, ``Low-cost multiple
  bit upset correction in sram-based fpga configuration frames,'' {\em IEEE
  Transactions on Very Large Scale Integration (VLSI) Systems}, vol.~24,
  pp.~932--943, March 2016.

\bibitem{keccak1}
G.~Bertoni, J.~Daemen, M.~Peeters, and G.~Van~Assche, {\em Keccak},
  pp.~313--314.
\newblock Berlin, Heidelberg: Springer Berlin Heidelberg, 2013.

\bibitem{5682391}
S.~P. Park, D.~Lee, and K.~Roy, ``Soft-error-resilient fpgas using built-in 2-d
  hamming product code,'' {\em IEEE Transactions on Very Large Scale
  Integration (VLSI) Systems}, vol.~20, pp.~248--256, Feb 2012.

\bibitem{6187474}
M.~Poolakkaparambil, J.~Mathew, A.~M. Jabir, and S.~P. Mohanty, ``Low
  complexity cross parity codes for multiple and random bit error correction,''
  in {\em Thirteenth International Symposium on Quality Electronic Design
  (ISQED)}, pp.~57--62, March 2012.

\bibitem{7167341}
R.~Santos, S.~Venkataraman, and A.~Kumar, ``Dynamically adaptive scrubbing
  mechanism for improved reliability in reconfigurable embedded systems,'' in
  {\em 2015 52nd ACM/EDAC/IEEE Design Automation Conference (DAC)}, pp.~1--6,
  June 2015.

\bibitem{4211783}
F.~Dittmann and S.~Frank, ``Hard real-time reconfiguration port scheduling,''
  in {\em 2007 Design, Automation Test in Europe Conference Exhibition},
  pp.~1--6, April 2007.

\bibitem{8351649}
M.~M. Wong, J.~Haj-Yahya, S.~Sau, and A.~Chattopadhyay, ``A new high throughput
  and area efficient sha-3 implementation,'' in {\em 2018 IEEE International
  Symposium on Circuits and Systems (ISCAS)}, pp.~1--5, May 2018.

\bibitem{Satish1450079}
N.~R. Satish, K.~Ravindran, and K.~Keutzer, ``Scheduling task dependence graphs
  with variable task execution times onto heterogeneous multiprocessors,'' in
  {\em Proceedings of the 8th ACM International Conference on Embedded
  Software}, EMSOFT '08, (New York, NY, USA), pp.~149--158, ACM, 2008.

\end{thebibliography}
}
\end{document}